%% file: main.tex
\documentclass[conference]{IEEEtran}
\IEEEoverridecommandlockouts
\usepackage{cite}
\usepackage{amsmath,amssymb,amsfonts}
\usepackage{algorithmic}
\usepackage{graphicx}
\usepackage{textcomp}
\usepackage{xcolor}
\usepackage{physics} % SFELD
\usepackage{subcaption} %MF
\usepackage{confcopyright} %ALANKHEIT
\def\BibTeX{{\rm B\kern-.05em{\sc i\kern-.025em b}\kern-.08em
    T\kern-.1667em\lower.7ex\hbox{E}\kern-.125emX}}
\begin{document}

\title{Optimizing Geometry Compression using Quantum Annealing\\
}

\author{\IEEEauthorblockN{Sebastian Feld, Markus Friedrich, Claudia Linnhoff-Popien}
	\IEEEauthorblockA{\textit{Institute for Computer Science} \\
		\textit{LMU Munich}\\
		Munich, Germany \\
		\{sebastian.feld, markus.friedrich, linnhoff\}@ifi.lmu.de}
}

\maketitle

\newcommand{\snot}[1]{\overline{#1}}

\begin{abstract}
The compression of geometry data is an important aspect of bandwidth-efficient data transfer for distributed 3d computer vision applications. 
We propose a quantum-enabled lossy 3d point cloud compression pipeline based on the constructive solid geometry (CSG) model representation.
Key parts of the pipeline are mapped to NP-complete problems for which an efficient Ising formulation suitable for the execution on a Quantum Annealer exists.
We describe existing Ising formulations for the maximum clique search problem and the smallest exact cover problem, both of which are important building blocks of the proposed compression pipeline.
Additionally, we discuss the properties of the overall pipeline regarding result optimality and described Ising formulations.
\end{abstract}


\IEEECopyright{https://ieeexplore.ieee.org/document/8644358}{10.1109/GLOCOMW.2018.8644358}{2018} %ALANKHEIT

\input{sections/intro.tex}

\input{sections/background.tex}

\input{sections/relatedwork.tex}

\input{sections/csg.tex}

\input{sections/quantum.tex}

\input{sections/discussion.tex}

\input{sections/conclusion.tex}

\bibliographystyle{alpha} % SFELD
\bibliography{main} % SFELD

\end{document}

%% file: sections/intro.tex
\section{Introduction}
\label{intro}

3d computer vision and data visualization represent two important topics of the digital era. 
Both deal with enormous amounts of (sensor) data that needs to be formalized, filtered, and transferred dependent on employed algorithms. 
The ultimate goal is to extract semantic knowledge either automatically (in case of computer vision) or with a human in the loop (data visualization). This abstract knowledge is required in a multitude of different domains, e.g. autonomous driving, 3d printing, augmented reality or virtual reality.

A widely used representation of real-world objects scanned by 3d sensors is the point cloud. 
Each contained element stores a sensor value represented as a 3d coordinate. Nowadays, a vast amount of 3d sensors is available in consumer hardware like smartphones or game consoles that complement professional high-accuracy systems. 
All of them are able to produce huge 3d data-sets. 
In order to get a more compact representation and a deeper understanding of a scanned real-world scene, more sophisticated geometry encodings are necessary. 
One of such encodings is the so-called \textbf{constructive solid geometry} (CSG) representation which approximates objects with a set of geometric primitives (e.g. spheres, boxes, or cylinders) combined with boolean operators (e.g. unions or intersections). 
CSG encoded geometry not only represents 3d data-sets in a more abstract way but also reduces necessary bandwidth for machine-to-machine communication. 
Among other problems in point cloud processing, such as point cloud registration and per-point normal estimation, the conversion from point cloud to CSG is difficult to solve efficiently. 

Quantum computing basically means to perform computation using quantum effects, whereas \textbf{quantum annealing} (QA) is a special case based on the adiabatic theorem. 
QA essentially performs classical simulated annealing but exploits quantum effects such as superposition, entanglement, and quantum tunneling. 
The idea of QA is that one can retrieve the global optimal solution of a real-world optimization problem once it has been transferred to a quadratic unconstrained binary optimization (QUBO) problem. 
The theory behind QA is rather old, but tangible results can be found only recently.

This paper discusses the two briefly outlined research areas, namely point cloud compression using CSG encoding and quantum annealing, and connects them accordingly. 
First we give a detailed description of the 3d point cloud to CSG conversion problem and how it can be split into subordinated optimization problems that can then be solved successively. Secondly, we discuss potential mappings of the above-mentioned optimization problems to well-known and precisely formulated NP-complete problems. 
There exists a wide range of reformulations from Karp’s NP-complete problems \cite{karp1972reducibility} to Ising models \cite{lucas2014ising}, that are the base for solving problems formulated as QUBOs. 
Finally, we summarize our investigations on how to solve the CSG conversion problem with help of the aforementioned QUBO formulations of selected NP-complete problems.
It offers first answers to the research question if -- and to what extent -- quantum annealing can be applied advantageously in selected problems of 3d geometry processing.

This paper makes the following contributions: (1) A design of a 3d point cloud compression pipeline that uses CSG trees as output representation and an Ising formulation of sub problems that can be solved on QA hardware. (2) A description of the Ising formulations of the maximum clique search problem and the smallest exact cover problem and their suitabality for the proposed pipeline. (3) A discussion of the proposed pipeline with respect to problem partitioning, resulting CSG tree optimality and described Ising formulations.

The remainder of this paper is structured as follows: 
Section \ref{background} introduces important concepts in the domains of CSG tree extraction and quantum annealing. 
Section \ref{relatedwork} offers an overview of existing point cloud compression methods and techniques for CSG tree extraction, and concludes with a description of real-world problems solvable with quantum annealing. 
Section \ref{csg} derives mappings of CSG extraction process steps to problems with existing Ising formulations. In addition, it details the extraction pipeline.
Section \ref{quantum} follows with a description of the aforementioned Ising formulations, while Section \ref{discussion} discusses the proposed approach. Finally, we conclude the paper in Section \ref{conclusion}.

%% file: sections/background.tex
\section{Background}
\label{background}

\subsection{CSG Trees and the Extraction Problem}

\begin{figure}[] % \begin{figure}[!htbp]
	\centerline{\includegraphics[width=0.31\textwidth]{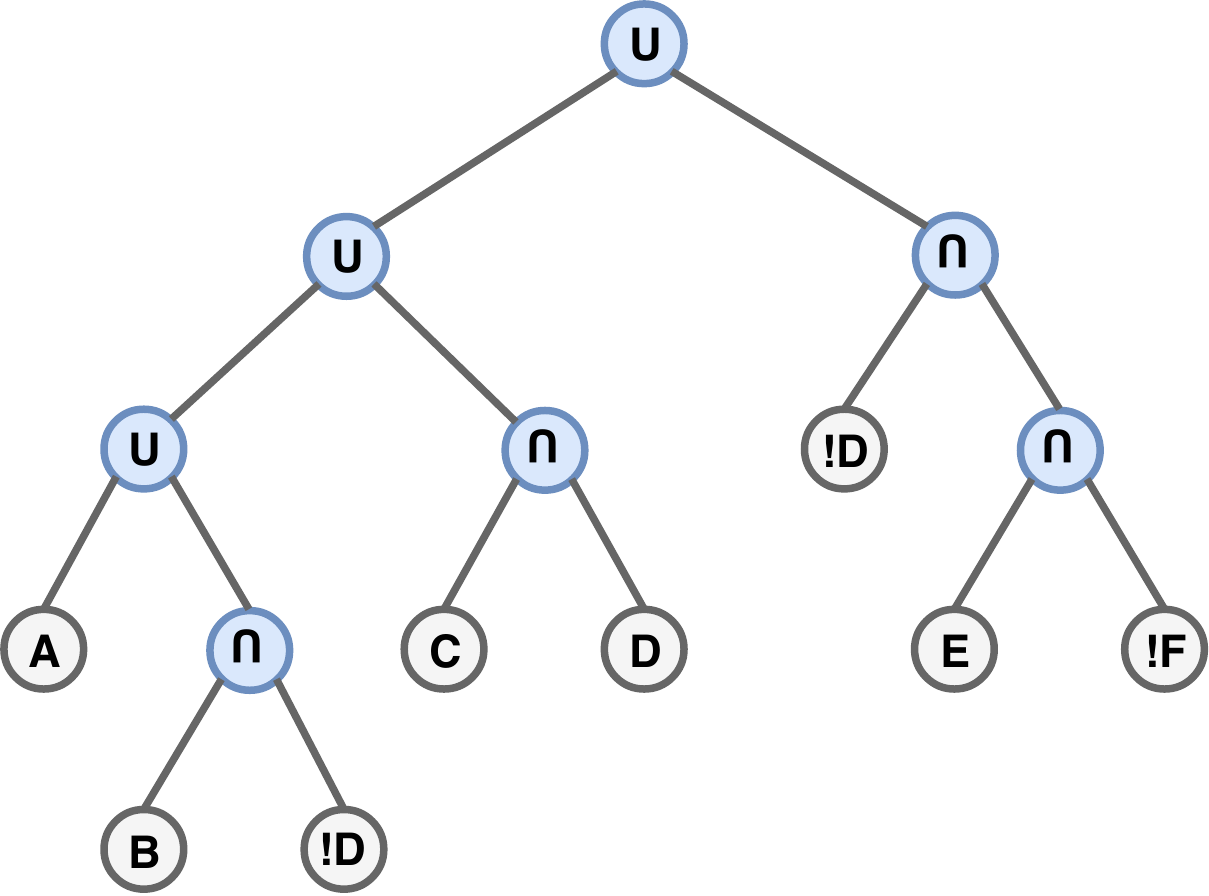}}
	\caption{Optimal CSG tree for surface $S$ denoted in Figure \ref{fig:example0}. The complement set operator is noted as ``$!$''.}
	\label{fig:csg_opt}
\end{figure}

The surface $S$ of a 3d model can be represented by a set of geometric primitives $P$ and a tree $\Phi(P)$ that contains these primitives in its leaves and Boolean set operations (union $\cup$, intersection $\cap$, and complement $\snot{\phantom{p}}$) as its inner nodes, see Figure \ref{fig:csg_opt} for an example. 
If the shapes of primitives $P$ are described implicitly as the zero sets of analytic signed distance functions $f_{p_{i}}: \mathbb{R}^3 \mapsto \mathbb{R}$, one can speak of $\Phi(P)$ being a semi-analytic representation of the 3d model.

The set $S=|\Phi(P)|$ describes the surface uniquely, however, more than one CSG tree might represent the same surface.    
Thus, an important aspect of the CSG tree representation is its size, i.e. number of leaves.
A CSG tree $\Phi$ for surface $S$ is absolutely minimal if there does not exist a smaller tree that represents $S$.
Finding the absolutely minimal tree for $S$ is in complexity class NP and equivalent to Boolean function simplification \cite{Lawler1964}. 

\begin{figure}[] % \begin{figure}[!htbp]
	\centerline{\includegraphics[width=0.4\textwidth]{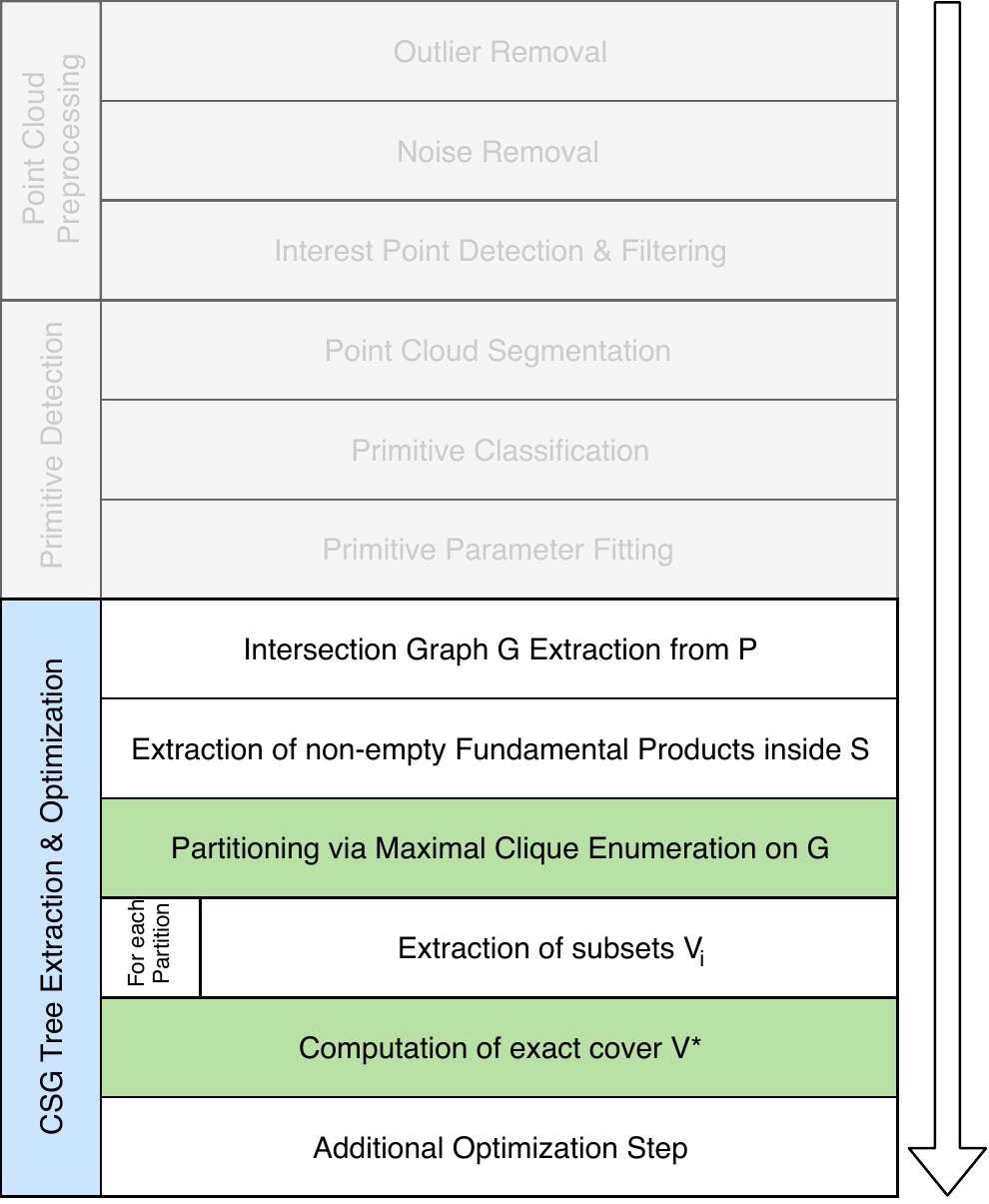}}
	\caption{The full CSG tree extraction pipeline. Point cloud pre-processing and primitive detection are not considered in this work (grey). The steps for the CSG tree extraction and optimization (blue) that are highlighted in green can be executed on quantum annealing hardware.}
	\label{fig:pipe}
\end{figure}

The considered CSG tree extraction problem tries to derive a CSG tree $\Phi(P)$ representation from a point cloud $C$ that represents the 3d model.
The extraction process follows a common pipeline model (see also Figure \ref{fig:pipe}): 
Noise and outliers are filtered from $C$. 
Then, primitives $P$ are segmented (which parts of $C$ are covered by $p_i$?), classified (what kind of primitive is $p_i$?) and fitted (what parameters describe $p_i$?). 
With known $P$, a $\Phi(P)$ representing $S$ is extracted and further minimized in a final optimization step.

For this paper, we focus on the last two steps, the extraction and minimization of $\Phi(P)$ for known $P$ and consider the previous steps as solved. 

\subsection{Quantum Annealing for NP-complete Problems }

% KOMBINATORISCHE OPTIMIERUNG (MA SEBASTIAN)
(Combinatorial) optimization problems are ubiquitous. A simple and demonstrative example is the optimization of a portfolio of stocks. Given stocks with estimated risks and opportunities together with a budget, the portfolio selection problems asks for spending the budget as completely as possible while minimizing the risk and maximizing the outcome \cite{goldfarb2003robust}. Further optimization problems can be found in the domain of logistics (capacitated vehicle routing problem \cite{ralphs2003capacitated}) or the distribution of chips on circuit paths \cite{hopfield1985neural}. Many optimization problems are NP-hard. According to the current state of knowledge, however, these problems cannot be solved in polynomial time using a deterministic machine, i.e., today's classic computer \cite{troyer2005computational}.

% ANSAETZE ZUR LOESUNG (MA SEBASTIAN)
Numerous different methods for solving such NP-hard problems have been developed. Examples are genetic algorithms \cite{de1989using}, neural networks \cite{anderson1988neural}, or dynamic programming \cite{woeginger2003exact}. If a decision problem is rewritten into an optimization problem, also tabu search \cite{glover1998tabu} or simulated annealing \cite{kirkpatrick1983optimization} may be utilized. In addition, there is a method called quantum annealing that uses the effects of quantum mechanics \cite{albash2016adiabatic}. The idea behind this approach is quite old \cite{apolloni1988numerical, apolloni1989quantum, finnila1994quantum, kadowaki1998quantum}, but only recently the possibility for the implementation in practice arose. In 2010, D-Wave Systems manufactured the first commercially available quantum annealing hardware \footnote{https://www.dwavesys.com/our-company/meet-d-wave}.

Quantum annealing (QA) is, similar to the better known simulated annealing (SA), a heuristic approach for solving combinatorial optimization problems that is based on natural processes in the real world \cite{mcgeoch2014adiabatic}.
Optimization problems can visually be represented as a hilly landscape that maps all possible solutions to a problem, with the lowest point being the global minimum and optimal solution. A heuristic such as QA now searches this landscape for the global optimum, and it may occur that the algorithm gets stuck in a local minimum. QA now has the special feature that a certain tunneling coefficient $\Gamma$ can be used to tunnel through hills, thus escaping a local minimum. The coefficient $\Gamma$ is initialized with a high value in order to supply more kinetic energy to the solution space in form of quantum fluctuations at the beginning of the annealing process. Over time, the value of $\Gamma$ decreases continuously, such that the algorithm can tunnel larger hills initially and approaches an optimum towards the end.

% ISING MODEL (MA CHRISTOPH)
The Ising model originates from theoretical physics and can be used to describe phase transitions and certain properties of particles in a system that evolves over time \cite{baxter2016exactly}. In particular, D-Wave's QA hardware is able to solve (optimization) problems mapped to the Ising model.
This model can be illustrated as a magnet consisting of $n$ molecules located on the nodes of a graph $G=(V,E)$. Each molecule is described by spin variables $s_i$ and can be in one of two configurations: either a parallel spin alignment regarding a particular axis (value +1) or anti-parallel (value -1) \cite{baxter2016exactly}. The energy of a spin configuration in such a model is defined as
\begin{equation}
H(s_1,...,s_n)= \sum \limits_{i}h_is_i+\sum \limits_{i<j}J_{ij}s_is_j
\end{equation}
where $h_i$ results from intermolecular forces within the magnet and $J_{ij}$ describes forces and interactions between the molecules. Many problems (and thus also optimization problems) that can be formulated as an Ising model additionally require to find the spin configuration with the lowest energy, i.e., to minimize the value of $H(s_1,...,s_n)$.

% QUBO (MA CHRISTOPH)
% Das Ising Model ist eng verwandt mit der Quadratic Unconstrained Binary Optimization (QUBO), welches anstelle von Spins, die einen Wert von $\pm1$ annehmen koennen, binaere Variablen $x_i \in \{0,1\}$ verwendet. Somit kann die QUBO ebenfalls als ein gewichteter Graph $G=(V,E,c)$ mit der Knotenmenge $V=\{x_1,v_2,...,x_n\}$ und der Kantenmenge $E=\{(i,j): i,j\in N\}$ mit den Kantengewichten $c_{ij}$ dargestellt werden. Das QUBO Problem wird dann folgendermassen definiert: $\min \sum \limits_{i\in N}c_{ij}x_i + \sum \limits_{(i,j)\in E}c_{ij}x_ix_j \qquad \text{mit } x_i = \{0,1\} \text{ und } i \in N$. Eine aequivalente Formulierung, bei der die Koeffizienten aus voriger Gleichung als Matrix $Q$ dargestellt werden, ist: $\min x^tQx \qquad \text{mit } x \in \{0,1\}^n$. Hierbei entspricht $Q$ einer $n\times n$ symmetrischen Koeffizienten-Matrix und $x$ einem Vektor bzw. $x^t$ dem transponierten Vektor \cite{lewis2017quadratic}.
% Aufgrund der erwaehnten Aehnlichkeit zum Ising Model, kann das QUBO einfach in das Ising Model ueberfuehrt werden und umgekehrt. Hierzu werden die Spin-Variablen im Ising Model durch $s_i=1-2x_i$ ersetzt und die Koeffizienten entsprechend angepasst. Die Energiefunktion des Ising Models hat somit die gleiche Form wie die Kostenfunktion der QUBO und laesst die Schlussfolgerung zu, dass beide Modelle aequivalent sind. Um nun ein bestimmtes Problem mit Hilfe eines Quantum Annealers loesen zu koennen, muss es lediglich auf solch eine QUBO Matrix bzw. Ising Model abgebildet werden.

%% file: sections/relatedwork.tex
\section{Related Work}
\label{relatedwork}

\subsection{Lossy Point Cloud Compression}

Lossy point cloud compression methods usually discretize point coordinates and store them in a hierarchical space partitioning structure \cite{Schnabel2006, Botsch2002, Peng2005}. 
Thus, loss is proportional to the size of the smallest partitioning cells.
Other techniques simply detect a set of 3d keypoints from the point cloud to be compressed \cite{Garstka2016,Hansch2014}. 
Of course, feasibility of keypoint description and detection techniques are dependent on the problem domain.
Geometric methods try to approximate point clouds with geometric primitives. 
For example, the authors of \cite{morell2014geometric} extract planes that roughly describe the shape of the input point-set.  
Our approach is similar as it extracts geometric primitives arranged in a CSG tree as a compressed representation of the input point cloud.

\subsection{CSG Tree Extraction}

CSG tree extraction and optimization from 3d models described through a set of quadrics was first investigated by Shapiro et al. (see \cite{Shapiro1991} for an overview of the problem and its solutions).
The approach by Fayolle et al. describes an extraction pipeline that works directly on the point cloud as input \cite{Fayolle2016}.
Geometric primitives are found and fitted using a RANSAC (Random Sample Consensus)-based method \cite{Schnabel2007}.
The extraction process is formulated as a combinatorial optimization problem (as described in Section \ref{ch:combi}) and solved using an evolutionary algorithm.
We combine a problem partitioning strategy with a CSG tree extraction method based on the smallest exact set cover problem that can be solved on quantum annealing hardware.

\subsection{Use Cases for Quantum Annealing}

There is a wide range of formulations of (theoretical) problems that can be solved using QA hardware \cite{lucas2014ising}. Numerous findings have been published that address specific applications. This includes research on wireless networking and scheduling \cite{wang2016quantum}, traffic flow optimization using real-world GPS data \cite{neukart2017traffic}, and many more.
Moreover, there are many studies on the concrete use of QA hardware in practice. For example, \cite{mcgeoch2013experimental, king2017quantum, king2015benchmarking} examine and benchmark how the solutions' quality of D-Wave's QA hardware behaves compared to those of classical optimization methods.
Finally, there are studies suggesting methods to further improve results of QA hardware \cite{king2014algorithm}.

%% file: sections/csg.tex
\section{The CSG Tree Extraction Problem}
\label{csg}

\subsection{CSG Extraction as a Combinatorial Optimization Problem}
\label{ch:combi}

With known primitives $P$, a CSG tree extraction process has to find the tree's topology (size, node and edge structure) and the correct assignment of primitives and operations to the tree nodes such that the result $\Phi(P)$ optimally approximates the target surface $S$.	 

A possible strategy defines a maximum tree size and formulates a combinatorial optimization problem over all possible combinations of tree topologies and node assignments together with an objective function that minimizes geometric error with respect to the input point cloud while penalizing large trees.
The combinatorial explosion can be mitigated by exploiting geometric relations between primitives, e.g. if two primitives do not intersect, they should not be operands of a boolean set operation in the tree. 
Dependent on the chosen solver, this approach is capable of producing absolutely minimal trees but at the cost of high computation times even for small ($\le 10$ primitives) trees \cite{Fayolle2016}. 

\subsection{CSG Tree Topology Constraints}

\begin{figure}[] % \begin{figure}[htb]
	\centering
	\begin{tabular}[c]{cc}
		\begin{subfigure}[c]{0.36\linewidth}
			\includegraphics[width=\textwidth]{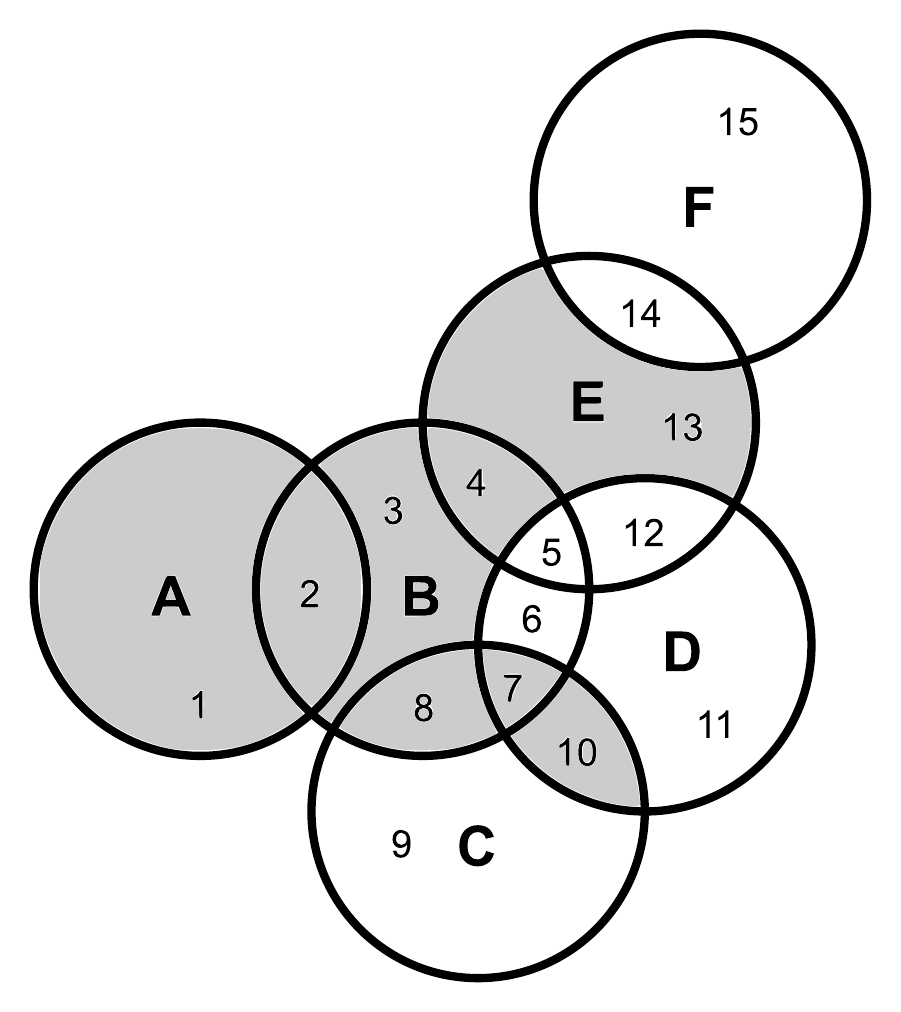}
			\caption{ }
			\label{fig:example0}
		\end{subfigure}&
		\begin{subfigure}[c]{0.36\linewidth}
			\includegraphics[width=\textwidth]{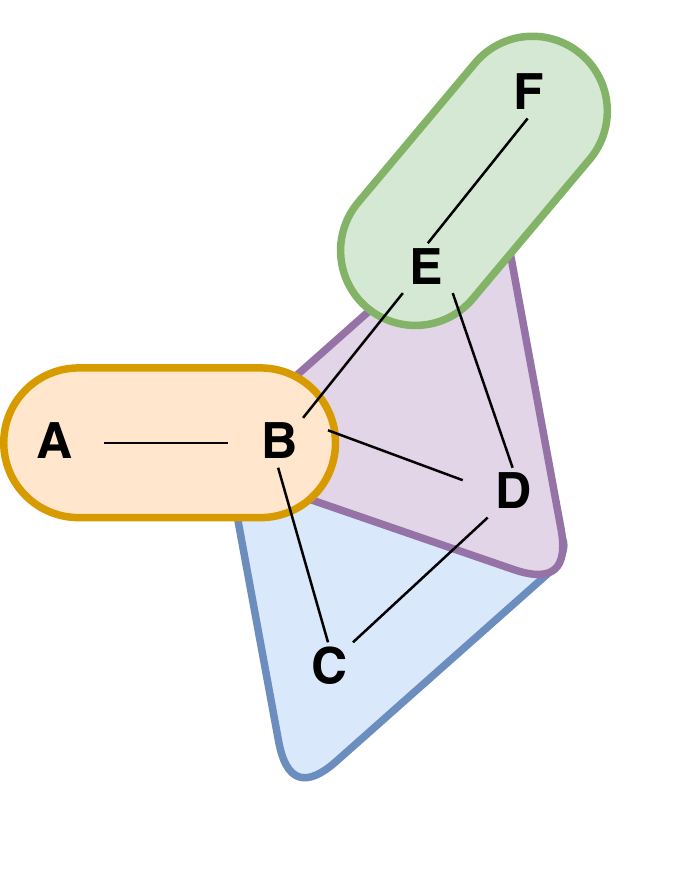}			
			\caption{ }
			\label{fig:example1}
		\end{subfigure}	
	\end{tabular}
	\caption{(a) Example geometry with $P=\{A,B,C,D,E,F\}$ and surface to represent $S$ in grey. Numbers $1$-$15$ identify non-empty fundamental products. (b) The corresponding intersection graph $G$ with the maximal clique set $Q=\{\{A,B\}, \{B,C,D\}, \{B,D,E\}, \{E,F\}\}$ outlined in black.}
\label{fig:example}
\end{figure}

Another solution to the combinatorial explosion is to define the topology beforehand.
Analogous to the disjunctive canonical form (DCF) for Boolean functions one can restrict the tree topology to a set of primitive intersections (so-called fundamental products \cite{Shapiro1991}) that are combined via the union set operator.
The result is commonly referred to as a two-level CSG representation \cite{Shapiro1991} and reads
\begin{equation}
\Phi(P) = \bigcup_{k=1}^{2^{|P|}-1} g_1 \cap g_2 \dots \cap g_{|P|}, g_i \in \{p_i, \overline{p_i}\}
\end{equation}
See Figure \ref{fig:example0} for an example geometry with labeled non-empty fundamental products.

The number of non-empty fundamental products $n_{f}$ is in $[|P|, 2^{|P|}-1]$.
It depends on the intersections between primitives which can be formalized using the intersection graph $G=(P,I)$ of the primitives $P$, where $I$ is the set of edges.
Each edge $(p_i, p_j)$ represents an intersection between primitive $p_i$ and $p_j$ with $i,j \in \{1,\dots, |P|\}$.
If $G$ is fully connected, $n_{f}$ reaches its maximum, if $I$ is empty its minimum.
See Figure \ref{fig:example1} for an examplary intersection graph.

This formulation reduces the aforementioned combinatorial optimization problem to a computational complexity of $\mathcal{O}(2^{|P|})$ since it only has to be checked for each non-empty fundamental product if it is inside the target surface $S$.
Its disadvantage is the size of the resulting tree and that even a minimal two-level CSG tree might not be the absolutely minimal tree representing $S$ \cite{Shapiro1991}.
The set $U$ of fundamental products inside $S$ for the example in Figure \ref{fig:example} is $\{1,2,3,4,7,8,10,13\}$.
The corresponding tree expression is $
(A\cap\snot{B})\cup 
(A\cap B) \cup 
(\snot{A} \cap B \cap \snot{C} \cap \snot{D} \cap \snot{E}) \cup 
(B \cap \snot{D} \cap E) \cup
(B \cap C \cap D) \cup 
(B \cap C \cap \snot{D}) \cup
(\snot{B} \cap C \cap D) \cup
(\snot{B} \cap \snot{D} \cap E \cap \snot{F}) 
$. 
In this case, the absolutely minimal tree would be $A \cup (B \cap \snot{D}) \cup (C \cap D) \cup (\snot{D} \cap E \cap \snot{F})$ which is $70\%$ smaller (see Figure \ref{fig:csg_opt} for the corresponding CSG tree).

\subsection{CSG Tree Size Optimization}

Now the question is how to improve this method for obtaining smaller CSG trees. 
A specific observation is of importance: Some parts of $S$ can be represented with less operations. 
In the example (Figure \ref{fig:example}), the part covered by fundamental products $\{7,10\}$ can also be covered with expression $(C\cap D)$ instead of $(B \cap C \cap D) \cup (\snot{B} \cap C \cap D)$.
Thus, if all possible expression combinations are considered, the optimal tree could be found (as stated in Sec. \ref{ch:combi}). 
Enumerating all possible expressions is equivalent to finding the set $V$ of all possible subsets of non-empty fundamental products that are inside $S$.
This is true since for each expression that represents a part of $S$, a set of fundamental products exists that represents the same part.

With set $V$ of fundamental product subsets, and set $U$ of fundamental products that represent $S$, it is possible to derive the minimal expression by finding the minimum exact cover $V^*$ which is a subset of $V$ such that each element of $U$ is covered by exactly one subset in $V^*$. 
The problem is NP-complete \cite{karp1972reducibility} but an Ising formulation exists \cite{lucas2014ising}. 
%\\

\subsection{Problem Partitioning}
\label{ch:part}

For the number of subsets $|V|$ to consider the upper bound 
\begin{equation}
\label{eq:part}
|V| \le \sum_{k=1}^{n_f}\binom{n_f}{k}= 2^{n_f}-1
\end{equation}
can be derived. 
In order to scale better with the number of considered primitives, a topological partitioning approach is applied.
It is based on the intersection graph $G$ which is partitioned into its maximal cliques $Q$.
We use this partitioning because it guarantees correctness of per-partition results, see Section \ref{discussion} for more details. 
Finding all maximal cliques in a graph is NP-hard \cite{eblen2012maximum} but a formulation of the related maximum clique search problem for the Ising model exists.

The partitioning has significant impact on the upper bound for $|V|$ which is now 
\begin{equation}
|V| \le \sum_{j=1}^{|Q|} 2^{n_f^j}-1.
\end{equation}
Thus, smaller cliques result in better scalability. 
Please note that resulting tree size is not optimal anymore:
For the example in Figure \ref{fig:example}, the size of the resulting tree increases slightly to $(A \cap \boldsymbol{\snot{B}}) \cup (B \cap \snot{D}) \cup (C \cap D) \cup (\boldsymbol{\snot{B}} \cap \snot{D} \cap E \cap \snot{F})$.
This effect is discussed in Section \ref{discussion}.

\subsection{Pipeline Overview}

The complete CSG tree extraction pipeline is depicted in Figure \ref{fig:pipe}. 

The CSG tree extraction starts with the generation of the intersection graph $G$ that is based on the parameters (position, orientation and geometry defining parameters, e.g. the radius of a sphere) of the primitives in $P$.
This step has $\mathcal{O}({|P|^2})$ computational complexity if implemented naively but can be improved to $\mathcal{O}(|P|\log({|P|} ))$ if hierarchical space partitioning schemes are used \cite{meagher1982Octree}.

Extraction of non-empty fundamental products $U$ is the next step. 
If $G$ is a fully connected graph, computational complexity is $\mathcal{O}({2^{|P|}})$.  
This step follows the problem partitioning where each partition is a maximal clique of $G$.
The Bron-Kerbosch algorithm is usually employed to solve the maximal clique enumeration problem and has a computational complexity of $\mathcal{O}((3.14)^{\frac{1}{3}|P|})$ in the worst case \cite{bron1973cliques}.
We propose to use an Ising formulation of the maximal clique enumeration problem. 

For each clique (partition) $q_i$ in $Q$, the set $V_i$ of possible subsets of $U_i$ is computed which has a computational complexity of $\mathcal{O}(3^{|q_i|})$ per clique. 
Note that now $V=\bigcup_{i=1}^{|Q|}V_i$.
Then $V^*$ being the smallest set of subsets from $V$ that exactly covers $U$ is computed.
The underlying smallest exact cover problem can be solved using the Dancing Links algorithm \cite{Knuth2000} with exponential computational complexity. 
We propose the use of an Ising formulation of this problem.

An additional optimization step tries to optimize the merged tree which is left for future work.

%% file: sections/quantum.tex
\section{Mapping of CSG-Tree Extraction Problems on Ising Models}
\label{quantum}

\subsection{Maximal Cliques Enumeration in Undirected Graphs}

Partitioning problems basically divide a set into two subsets. Given an undirected graph $G=(V,E)$, a clique $W \subseteq V$ is a subset of the vertices of $G$ forming a complete subgraph, i.e., any two vertices of $W$ are connected by an edge in $G$. The clique size $K=|W|$ is the number of vertices in $W$.

The question of whether or not a given graph contains a clique of size $K$ is an NP-complete decision problem \cite{karp1972reducibility}. There are formulations of problem Hamiltonians that solve this problem using $N$ logical qubits \cite{lucas2014ising} or, under certain circumstances, with slightly fewer logical qubits \cite{childs2000finding}.

Furthermore, there is the definition of a maximal clique being a clique that cannot be extended by incorporating another vertex. With other words, a maximal clique cannot be a subset of a larger clique. Lastly, finding a clique with a maximum number of vertices is called the maximum clique problem \cite{balas1986finding}. Again, there are several formulations for problem Hamiltonians that can be used to find (one of) the largest clique in a graph \cite{childs2000finding, lucas2014ising}. In some works the equivalence between the maximum clique problem and the maximum independent set problem is exploited \cite{boothby2016fast}.

\subsection{Smallest Exact Cover}

Karp's NP-complete problems contain several covering and packing problems, that have in common, that the constraints must exactly be satisfied.
Such problems are widely discussed in the adiabatic quantum optimization community because of their immense coverage on practical use and because these problems can easily be embedded on a graph that is not complete (like current QA hardware).

The exact cover can be described as follows \cite{lucas2014ising, choi2010adiabatic}: given a set $U = \{1,...,n\}$ with $n$ elements and given a set $\{V_i\}$ consisting of $N$ subsets of $U$, i.e. $V_i \subseteq U$ with $i=1,...,N$, such that the union of $\{V_i\}$ results in $U$, i.e. $U = \bigcup_{i=1}^{N} V_i$.
The question is now: Is there a subset of the set of subsets, i.e. $R \subseteq \{V_i\}$, such that the elements of $R$ are disjoint, i.e. $V_i \cap V_j = \emptyset$ for $i \neq j$, and such that the union of the elements of $R$ produce $U$, i.e. $\bigcup_{V_i \in R} V_i = U$? If yes, then $R$ is the exact cover.
An example is (see \cite{choi2010adiabatic}): $U=\{1,2,3,4,5\}$ with $n=5$, $\{V_i\}=\{V_1, ..., V_7\}$ with $N=7$, and $V_1=\{1,2,4\}$, $V_2=\{1,2,5\}$, $V_3=\{1,3,4\}$, $V_4=\{2,3\}$, $V_5=\{3\}$, $V_6=\{4,5\}$, $V_7=\{5\}$. The exact cover is $R=\{V_1,V_5,V_7\}$, since $V_1 \cup V_5 \cup V_7 = \{1,2,4\} \cup \{3\} \cup \{5\} = \{1,2,3,4,5\} = U$. 

Since there may be multiple solutions to the exact cover problem, one can also look for the smallest exact cover being set $R$ with as few elements as possible. \cite{lucas2014ising} formulate the problem Hamiltonian for the smallest exact cover problem as

\begin{equation}\label{eq:sec}
H = A \sum\limits_{\alpha=1}^n \left(1 - \sum\limits_{i:\alpha \in V_i} x_i\right)^2 + B \sum\limits_{i} x_i.
\end{equation}

The first term refers to the exact cover problem, the second term extends it to find the smallest exact cover. $\alpha$ denotes the elements of $U$, thus the first sum iterates over all $n$ elements of $U$. $i$ denotes the elements of $\{V_i\}$, i.e. the subsets $V_i$ of $U$. Thus, the second sum iterates over all $N$ elements of $\{V_i\}$. $x_i=1$ applies if the current element $\alpha \in U$ is also an element of the current element $V_i$, i.e. if $\alpha \in V_i$. It follows, that the term in the squared parentheses will have its lowest value of $0$ exactly if the element $\alpha \in U$ is included in exactly one element of $\{V_i\}$ ($\left(1-1\right)^2=0$). If the element $\alpha \in U$ is covered by none of the elements of $\{V_i\}$, we get $\left(1-0\right)^2=1$, and if it is covered by two (or more) elements, we get $\left(1-2\right)^2=1$. Summarized, the first term of equation \ref{eq:sec} will get the value of $0$ exactly if every element of $U$ is included exactly one time implying that the union of subsets of $\{V_i\}$ are disjoint. The existence of a value of $0$ for the first term implies the existence of a solution for the exact cover problem, while multiple possible assignments may occur.
The second term extends the exact cover problem to the smallest exact cover problem. Basically, it punishes the use of elements of $\{V_i\}$, meaning it favors solutions that require fewer elements $V_i$. The authors of \cite{lucas2014ising} suggest to use a ratio of $A > nB$ for a correct problem encoding. The problem Hamiltonian presented in equation \ref{eq:sec} can be realized using $N$ logical qubits.

%% file: sections/discussion.tex
\section{Discussion}
\label{discussion}

\subsection{Clique Partitioning and sub-optimal CSG Trees}

The reason for using a clique-based partitioning lies in the guaranteed correctness of per-clique tree results. 
As an example serves clique $\{A,B\}$ in Figure \ref{fig:example0}: 
In order to cover $S$ in the region influenced by primitive $A$ and $B$, both have to be combined via union.
However, correct and optimal merging is not trivial. 
A simple union of per-clique trees is only valid if primitives appearing in more than one clique are cut into fundamental products.
Otherwise, operations of one per-clique tree might be cancelled out by the merge union with another per-clique tree.
See example clique $\{A,B\}$ in Figure \ref{fig:example0}: Fundamental products $5$ and $6$ are not in $S$ but a simple union with the corresponding per-clique tree $A\cup B$ would erroneously cover them.

Cutting primitives enlarges the resulting tree which prevents optimality (see example in Section \ref{ch:part}).
Other per-clique merge strategies exist \cite{Friedrich2018} but more research is needed in both, partitioning schemes and merging strategies.

\subsection{Ising Formulations}

The Ising formulation of the smallest exact cover problem uses $1$ qubit per subset \cite{lucas2014ising}. 
Thus, on recent QA hardware with approximately $2000$ qubits, one can solve the extraction problem with $3$ primitives at most (see Equation \ref{eq:part}). 

For larger problem sizes, a partitioning scheme based on maximal clique enumeration is suitable for which, to the best of our knowledge, no Ising formulation exists. 
The Ising formulation of the related maximum clique search problem serves as a basis for future research. 
Another potentially suitable partitioning scheme is core-halo partitioning \cite{djidjev2016}.

%% file: sections/conclusion.tex
\section{Conclusion}
\label{conclusion}

We have proposed and discussed a lossy compression pipeline for point cloud data based on the CSG tree geometry representation that can profit from quantum annealing hardware. 
Open for future work is an implementation and thorough evaluation of the described approach.